\documentclass[fleqn,10pt]{wlscirep}
\usepackage{graphicx, listings}
\usepackage{epstopdf}
\usepackage{bbold}
\usepackage{wrapfig}

\newcommand{\ket}[1]{| #1 \rangle}

\newcommand{\Fref}[1]{Fig.~\ref{#1}}

\newcommand{\prl}{\emph{Phys. Rev. Lett.}}
\newcommand{\pra}{\emph{Phys. Rev. A}}
\newcommand{\rmp}{\emph{Rev. Mod. Phys.}}
\newcommand{\njp}{\emph{New J. Phys.}}
\newcommand{\natphys}{\emph{Nat. Phys.}}

\title{Thermometry of bosonic mixtures in Optical Lattices via Demixing}

\author[1,*]{F. Lingua}
\author[2]{B. Capogrosso-Sansone}
\author[3,4]{F. Minardi}
\author[1]{V. Penna}
\affil[1]{Department of Applied Science and Technology and u.d.r. CNISM, Politecnico di Torino, I-10129 Torino, Italy}
\affil[2]{Department of Physics, Clark University, Worcester, Massachusetts 01610}
\affil[3]{Istituto Nazionale di Ottica (INO-CNR)}
\affil[4]{European Laboratory for Non-Linear Spectroscopy (LENS) and Dipartimento di Fisica, Universit\`a di Firenze, I-50019 Sesto Fiorentino - Firenze, Italy}

\affil[*]{fabio.lingua@polito.it}

\begin{abstract}
  Motivated by recent experiments and theoretical investigations on binary
  mixtures, we investigate the miscible-immiscible transition
  at finite temperature by means of Quantum
  Monte Carlo. Based on the observation that the segregated phase is strongly affected by
  temperature, we propose to use the degree of demixing for thermometry of a
  binary bosonic mixture trapped in an optical lattice.  We show that the
  proposed method is especially sensitive at low temperatures, of the order of
  the tunnelling amplitude, and therefore is particularly suitable in the regime
  where quantum magnetism is expected.
\end{abstract}
\begin{document}

\flushbottom
\maketitle
\thispagestyle{empty}


Ultracold atoms in optical lattices are generally regarded as an almost ideal
experimental setting to investigate many-body quantum physics in strongly
correlated regimes \cite{2008-Bloch-ManyBodyReview}.
However, next to undisputed strengths, atomic systems suffer from some notable
limitations. Perhaps surprisingly if compared to condensed-matter counterparts,
for strongly correlated quantum gases measuring fundamental
parameters, such as temperature, is far from trivial, a fact often
encumbering the comparison between theoretical and experimental findings. The
physical reason is that the primary thermometric quantity used in cold atoms
experiments, namely the momentum distribution, in strongly correlated regimes is
often dominated by quantum rather than thermal fluctuations, thereby becoming
quite insensitive to temperature variations.  In recognition of its importance,
thermometry for optical lattices have sparked numerous theoretical proposals and
experiments \cite{2011-DeMarco-CoolingThermometryReview}. Quite generally, two
main approaches have been pursued for thermometry: through ancillary samples (or
sample subsets) in a well-understood, e.g. weakly interacting (WI), regime
\cite{2009-Ketterle-magnetism,2009-Ketterle-magnetism2,2010-Salomon-ThermodynamicsUFG,2010-DeMarco-SpinDependentOL,2016-Widera-ImpurityThermometry}; or by measuring {\em in-situ} local density
fluctuations with high-resolution imaging
\cite{2009-Greiner-microscope,Bloch2010,bakr2015,greiner2015,kuhr2015,cheuk2016}.
Other proposed methods still await experimental demonstration \cite{Ruostekoski2009,Zhou2011,Roscilde2014}.
\bigskip

\noindent
\textbf{A new thermometric scheme.} In this work, we propose a thermometry technique for ultracold quantum mixtures
in optical lattices based on the demixing of two mutually repulsive components.
Multi-components BECs have been created long ago \cite{Myatt1997,Stenger1998},
 while recently seminal works in thermometry have been reported both with bosonic and
fermionic samples. With a rubidium condensate, demixing between two spin
components was induced by a magnetic field gradient and the width of the
interface region was used to estimate the temperature
\cite{2009-Ketterle-magnetism,2009-Ketterle-magnetism2}. Recently, the spin waves, or `magnons', in a
spinor Rb condensate were used to reduce the entropy per particle to values as
low as $0.02\,k_B$ \cite{2015-StamperKurn-LowEntropy}, an order of magnitude below the values
required for the onset of magnetic phases \cite{2010-CapogrossoSansone-Entropy}.
On the fermionic side, anti-ferromagnetic correlations have been detected by means of Bragg
scattering \cite{2015-Hulet-BraggAFM}, and also, very recently, directly by means of single-site imaging \cite{2016-Greiner-AFM,2016-Zwierlein-AFM,2016-Kohl-AFM}.

We analyze the effect of temperature fluctuations on a demixed phase of a
mixture of two bosonic species trapped in an optical lattice with and without an external harmonic confinement. Demixing, i.e. the
spatial separation of the two components, occurs when inter-species repulsion
overcomes the intra-species one
\cite{1998-Timmermans-PhaseSeparation,1998-Chui-SegregatedPhase,pollet2013,indi_2016}. An important role is played by temperature
fluctuations which compete with, and eventually destroy, demixing \cite{1stpap}.
Here, we take advantage of this competition to propose a route for thermometry
in ultracold mixtures: we employ a suitable global estimator of demixing which
can readily be measured and used to determine the temperature.

We first consider the case of a moderately shallow optical lattice, where each species is
superfluid. At sufficiently weak interspecies interaction,
this regime is interesting as the momentum distribution of either species allows
an independent determination of the temperature \cite{2008-Trivedi-Thermal_nk} to
validate the currently proposed thermometry.
In the case of a deeper optical lattice, thus a strongly-interacting (SI) regime \cite{Gerbier_pra72},
we lack reliable temperature estimators other than the direct microscopic
observation of particle-hole pairs. In this regime, our proposed thermometry proves
a remarkably effective tool.
Specifically, we show how the degree of demixing characterizing the spatial
distribution of trapped mixtures depends upon the temperature.
In particular, demixing in low density regions is more susceptible to temperature fluctuations than in high density regions.

In this paper, we also probe the effect of temperature on spatial distribution,
showing that temperature-induced changes in the shell-structure of the trapped density correspond to a dramatic signature in the
boson interference patterns produced when the confining potential is turned off. Our results are based on large-scale path-integral quantum Monte Carlo simulations by a two-worm algorithm~\cite{2worm}.
\medskip

\begin{figure}[h!]
\begin{center}
  \includegraphics[width=0.7\columnwidth]{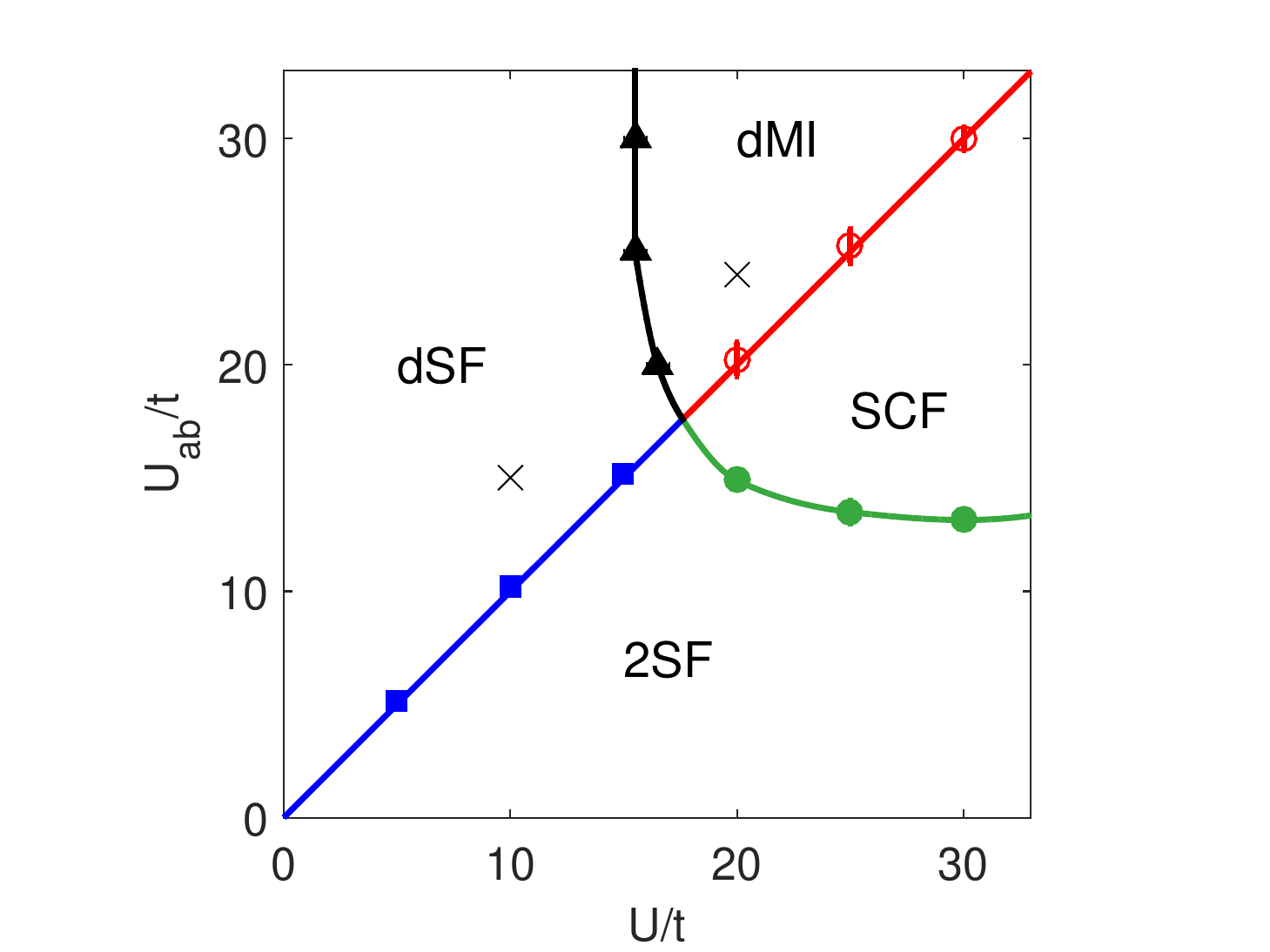}
  \caption{Ground-state phase diagram of twin bosonic species at total filling $n=1$, from
    \cite{1stpap}. Crosses indicate the parameters used here for simulations. The dSF (dMI) phase represents two spatially separated superfluids (Mott-Insulators). 2SF describes two mixed superfluids, and SCF is a global Mott-insulator phase where mobility is allowed as a superflow in the particle-hole channel.}\label{fig:PhaseDiagram}
\end{center}
\end{figure}

\noindent
\textbf{The Model and the phase diagram.} A mixture of two {\em bosonic} species trapped in a two-dimensional square
optical lattice, is described by the two-component Bose-Hubbard (BH) model:
\begin{equation}
H = H_a + H_b + U_{ab} \sum_{i} n_{ai}n_{bi}
\label{H1}
\end{equation}
where $U_{ab}$ is the inter-species repulsion, $n_{ai}$, $n_{bi}$ are the number
operators at site $i$ for species A and B respectively, and
\begin{equation}
H_c = \frac{U_c}{2} \sum_{i}n _{ci}(n_{ci} - 1) - t_c\sum_{\langle ij\rangle} c_{i}^{\dag}c_{j}
- \sum_{i}\mu_{ci}n_{ci}
\; ,
\label{Hc}
\end{equation}
where ${\langle ij\rangle}$ denotes to sum over nearest neighboring sites, $c=
a, b$ the bosonic species, $c_i$ ($c_i^\dag$) the annihilation (creation) operators
satisfying $[c_i, c_i^\dag]= 1$, $U_c$ the intra-species repulsion, $t_c$ the
hopping amplitude, and $\mu_{ci}=\mu_c$ ($\mu_{ci}=\mu_c - \omega_H {{\vec
    r}_{i}}^{\,\,2}$) the chemical potentials in the homogeneous (trapped) case.
In this paper, for simplicity, we consider a model with twin species, namely, we set $U_a = U_b= U$, $t_a = t_b= t$ and $\mu_a=\mu_b=\mu$. The condition on the chemical potentials implies that $N_a=N_b$, where $N_a$ and $N_b$ are the total number of particles of species $A$ and $B$, respectively.
The ground-state phase diagram
of a twin-species mixture at total integer filling features
a demixed superfluid (dSF), or a demixed Mott-insulator (dMI), when the interspecies
interaction becomes greater than the intraspecies repulsion, and a
double-superfluid phase (2SF) or a supercounterflow (SCF) otherwise. This is illustrated in
\Fref{fig:PhaseDiagram} where integer total filling factor $n=1$ has been assumed.

\section*{Results}

We first consider a homogeneous system of linear size $L$ (in unit of the lattice step $a$ which we set as our unit
length) with periodic boundary conditions. We work at integer total filling factor considering both $n=1$ and $n=2$.
This corresponds to the condition $N_a + N_b=nL^2$. Then we move to analyse a trapped system by introducing a harmonic
trapping potential.

To study the effect of temperature fluctuations, we
first focus on the transition 2SF-dSF since this can be observed for any choice
of the filling factor~\cite{1stpap}, while the transition from dMI-SCF requires
an integer value of the filling factor.

We quantify demixing effects through the parameter
\begin{equation}
\Delta=\frac{1}{M}\sum_i\Big[\frac{\langle n_{ai} \rangle -
\langle n_{bi}\rangle}{\langle n_{ai}\rangle + \langle n_{bi}\rangle}\Big]^2, \label{eq:Delta}
\end{equation}
where the sum runs over the $M=L^2$ lattice sites;
evidently, $\Delta$
ranges from 0, if all sites are equally populated, to 1, for complete demixing.


\subsection*{Homogeneous case}
\begin{figure}[h!]
\includegraphics[width=\textwidth]{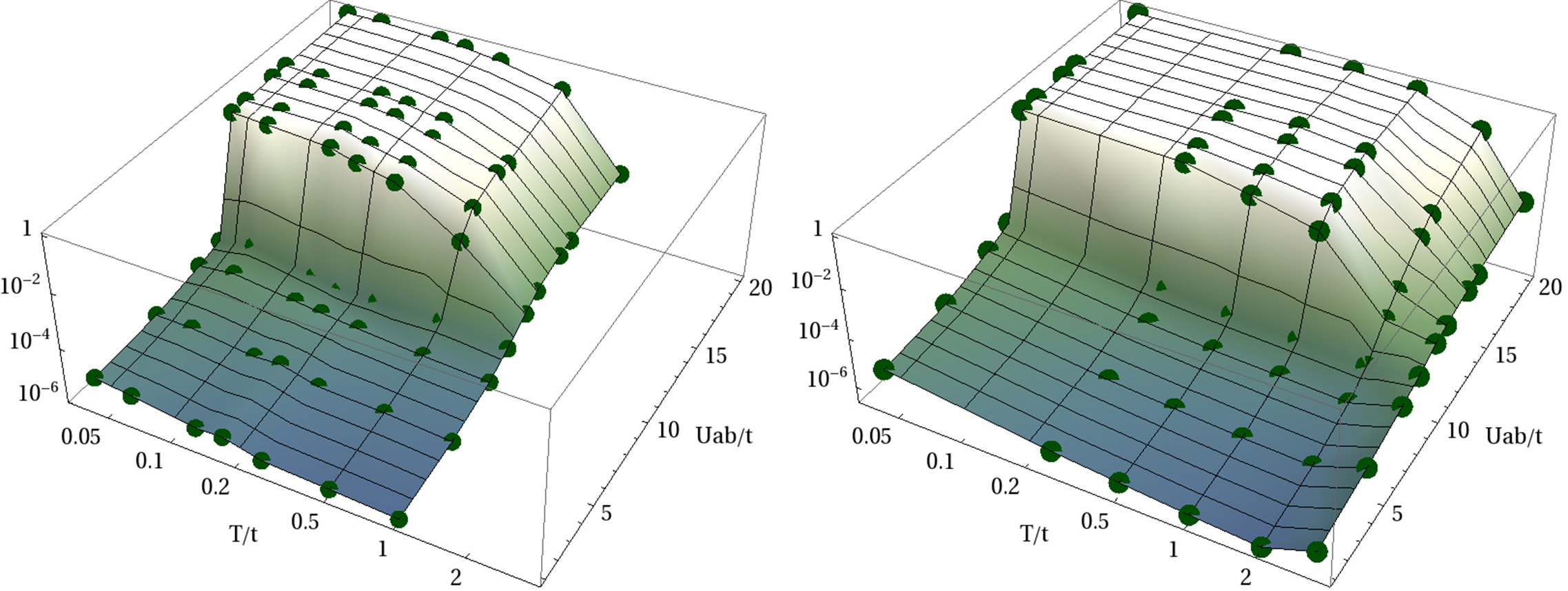}
\caption{Parameter $\Delta$ as a
    function of temperature $T/t$ and interspecies interactions $U_{ab}/t$, at $U/t=10$. Total filling
    $n=1$ (left panel) and $n=2$ (right panel). Points show numerical results.}
\label{fig:Delta_n1_n2}
\end{figure}

\Fref{fig:Delta_n1_n2} shows $\Delta$ as a function of temperature $T$ and
inter-species interactions $U_{ab}$, at fixed $U=10t$, for total filling $n=1$
(upper panel) and $n=2$ (lower panel). At lower temperatures, a step-like
increase in the value of the demixing parameter $\Delta$ signals the onset of
strong demixing in the system. As the temperature is increased, thermal
fluctuations become more prominent and the mixing of the two components is
restored even for $U_{ab}>U$.
We observe a pronounced dependence of $\Delta$ on temperature in the dSF phase, with a three order of
magnitudes drop within a range of temperatures of the order of the tunneling
energy $t$.  On the contrary, in the 2SF phase, $\Delta$ is rather insensitive
to the temperature, and it remains orders of magnitude smaller than in the dSF
phase.  As outlined below, the strong dependence of $\Delta$ on $T$ displayed in the dSF phase
motivates the basic idea of extracting the temperature from the measurement of the
demixing parameter.

For total filling factor $n=2$ (lower panel in \Fref{fig:Delta_n1_n2}), the
larger energy penalty associated with the overlap between components leads to a
more evident demixing \cite{1stpap}. Robustness against miscibility results in a stronger
robustness against temperature fluctuations than at lower filling
factors: notice the evident increase in temperatures needed in order to
destroy demixing for $n=2$.  In this sense, larger filling factors shift the
operating range of the proposed thermometer towards higher temperatures.
This behavior can be understood by observing that, for $U_{ab} > U$, increasing
the filling factor further inhibits the start of the mixing process of the two
components. In this sense, the minimal mixing consists in displacing a single boson $a$ ($b$) in the
domain of components B (A). For $n=2$, the heuristic calculation
of the free-energy cost for creating a double pair $ab$
gives $\Delta F = 2(U_{ab} -U) (n-1) -T k_B {\rm ln} (L^4/4)$.
The mixing temperature is found to be
$k_B T= 2(U_{ab} -U)/{\rm ln} (L^4/4)$.
For $n=1$, the mixing process begins with the formation of a single pair $ab$ and a hole. This entails
$\Delta F = U_{ab} -T k_B {\rm ln} (L^4/4)$ and $k_B T = U_{ab}/{\rm ln} (L^4/4)$.
In both cases, using the parameter values of \Fref{fig:Delta_n1_n2}
gives $k_B T/t \sim 1$
in agreement with numerical results.
The dependence on the lattice size $L$ reflects the finite-size character
of our model. The temperature at which the first
pairs $ab$ crop up is proportional to $U_{ab}$, thus confirming the
inhibition of the mixing effect for increasing $U_{ab}$.


\subsection*{Trapped system}
In order to consider a more general and realistic scenario, we
relax the homogeneity assumption
and study the system in a harmonic trap.  
The chemical potential of species $c=a,b$, transforms according to
\begin{equation}
\mu_{ci}=\mu_{c} - \omega_H^2 {\vec r}_i^{\,2}
\end{equation}
where $\omega_H$ is the curvature of the harmonic trap and ${\vec r}_i$ the position vector of lattice site
$i$.  This leads to a site-dependent filling factor $n_i$. Generally, demixing is not affected by the presence of a harmonic potential as far as the condition $U_{ab}>U$ is satisfied. Demixing in the trap manifests itself through the occurrence of a sharp and straight
boundary between the two species. This represents the minimum-energy
configuration for a demixed system in a trap, as originally predicted for continuous systems \cite{1998-Chui-SegregatedPhase}.
\begin{figure*}[ht!]
  \includegraphics[width=\textwidth]{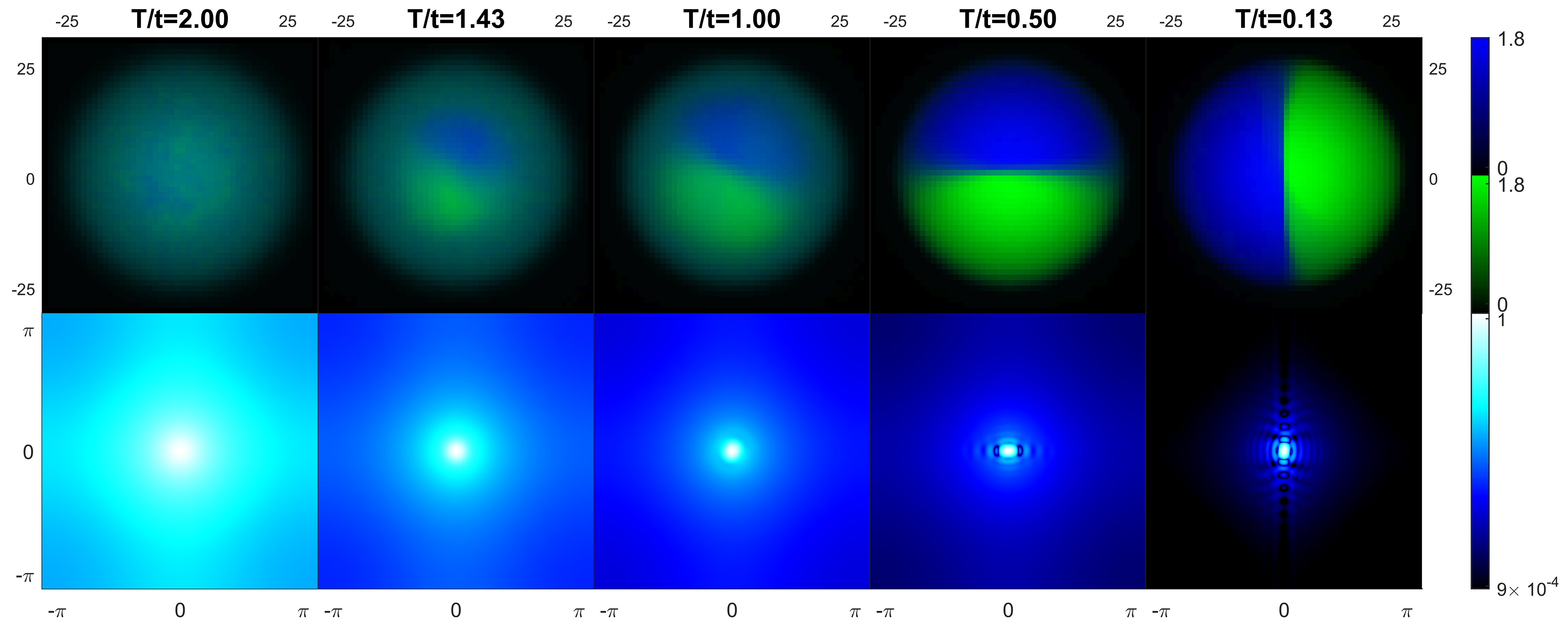}\caption{Density maps (first
    row) of species A (green) and B (blue), and computed momentum distributions of species B (second row) for
    decreasing temperature (left to right). The mixture ($U/t=10$,
    $U_{ab}/t=15$) is trapped in a harmonic potential of strength
    $\omega_H/t=0.03$.  }\label{fig:density_TOF_2Dplots}
\end{figure*}
\medskip

\noindent
\textbf{Weak Interaction.} In the following, we present finite temperature results for the trapped case of WI mixtures.  We find that a spatial shell structure arises at intermediate temperatures (see Fig. \ref{fig:density_TOF_2Dplots}), in which a central demixed phase (dSF)
is surrounded by a shell of mixed phase (2SF). In
the first row of \Fref{fig:density_TOF_2Dplots}, it is well visible that
the temperature-induced mixing effect first appears where the density is lower (that is in the outer shell)
and in the proximity of the boundary separating the two species.
Such an effect is due to the larger entropy associated with demixing  in these regions.
For sufficiently large temperatures we detected the presence of a third surrounding shell
of a Normal-Fluid (NF) phase.
As expected, the thickness of the NF shell increases for increasing temperatures (see Supplementary information for details).
\begin{figure}[h!]
\begin{center}
  \includegraphics[width=0.7\columnwidth]{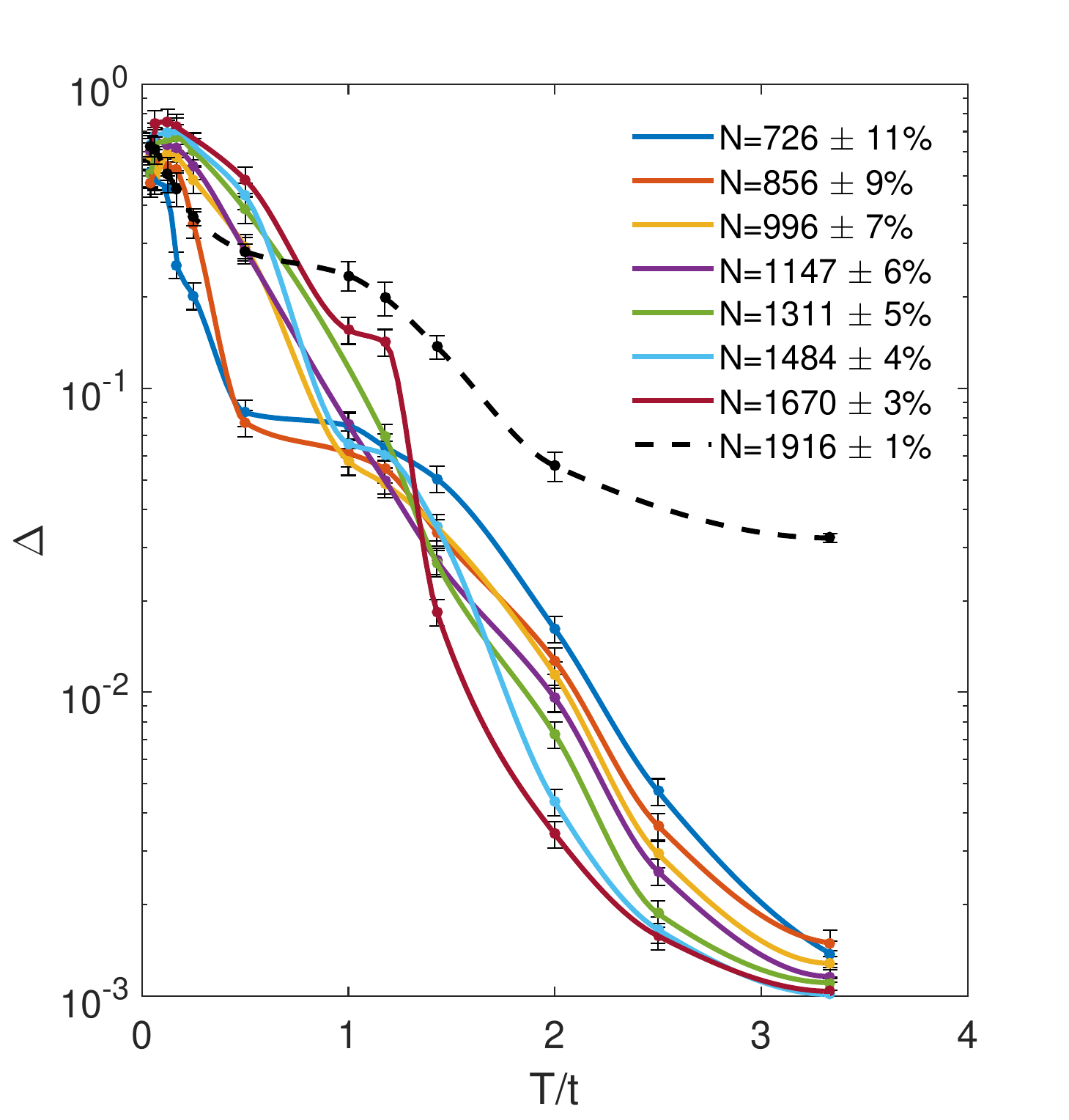}
\caption{$\Delta$ parameter as a
    function of temperature in the trapped case. Solid lines: $\omega_H/t=0.03, U/t=10, U_{ab}/t=15$; black dashed line:
 and $\omega_H/t=0.12, U/t=20, U_{ab}/t=24$.}
\label{fig:DeltaVsT_trap}
\end{center}
\end{figure}

In Fig. \ref{fig:DeltaVsT_trap} we show the behaviour of parameter $\Delta$
as a function of $T/t$ for the trapped case. We have performed
simulations for different number of atoms $N$, both in the
WI (solid lines) and  SI (black dashed line) regimes.
In the WI regime with $U/t=10$ and $U_{ab}/t=15$, besides confirming that a system with a higher number of bosons
is more robust against temperature mixing, we notice that, over the considered
range of temperature, the demixing parameter drops approximately one
order of magnitude less than in the homogeneous case. Larger local densities at
the center of the trap imply larger interactions energies thus reducing the boson
mobility and their mixing degree.

In search of additional experimental signatures
of the temperature-driven transition from the dSF to the 2SF phase,
 we computed the momentum distributions $n_{c,
  \mathbf{k}}=|\phi_{(\mathbf{k})}|^2 \sum_{i,j}e^{i\mathbf{k}(\mathbf{r}_i -
  \mathbf{r}_j)}\langle c^{\dag}_i c_j\rangle $ \cite{tof1} for species $c=a,b$.
Distributions $n_{c, \mathbf{k}}$, integrated along one direction, are recorded by the time-of-flight images
experimentally observed, provided that interactions have a negligible effect
during the expansion. In the second row of
Fig. \ref{fig:density_TOF_2Dplots} we plot the ${\mathbf k}$-periodic part of
momentum distribution, i.e.  $\tilde{n}_c({\mathbf k})\equiv n_c({\mathbf
  k})/|\phi_{(\mathbf{k})}|^2$, for species B (due to the symmetry of the system the momentum distributions of species A show the same features).
For increasing temperatures (right to left), we observe how the appearance of the spatial shell structure is accompanied by changes in
the momentum distribution.
When the ``hard-wall'' separating the demixed species
is present, the phase coherence of each species is restricted to the
portions of the lattice where the species is confined, and this produces the
fringes
shown in the $\tilde n_c({\mathbf k})$ images of Fig.
\ref{fig:density_TOF_2Dplots}. 
Such fringes arise due to the interference of waves bouncing back
from the ``hard-wall'' separating the demixed species.
\medskip

\noindent
\textbf{Strong Interaction.} After showing that parameter $\Delta$ is a convenient temperature
indicator for WI mixtures, we move to consider the case of
SI regimes. In \Fref{fig:DeltaVsT_trap} we plot the parameter
$\Delta$ as a function of temperature for larger interactions,
i.e. $U/t=20$, $U_{ab}/t=24$, and $\omega_H/t=0.12$.  We find that $\Delta$ still decreases as the temperature increases.
Due to the larger value of $U_{ab}/t$, $\Delta$ drops by at least one order of magnitude less than in the WI case.

\begin{figure*}[ht!]
  \includegraphics[width=\textwidth]{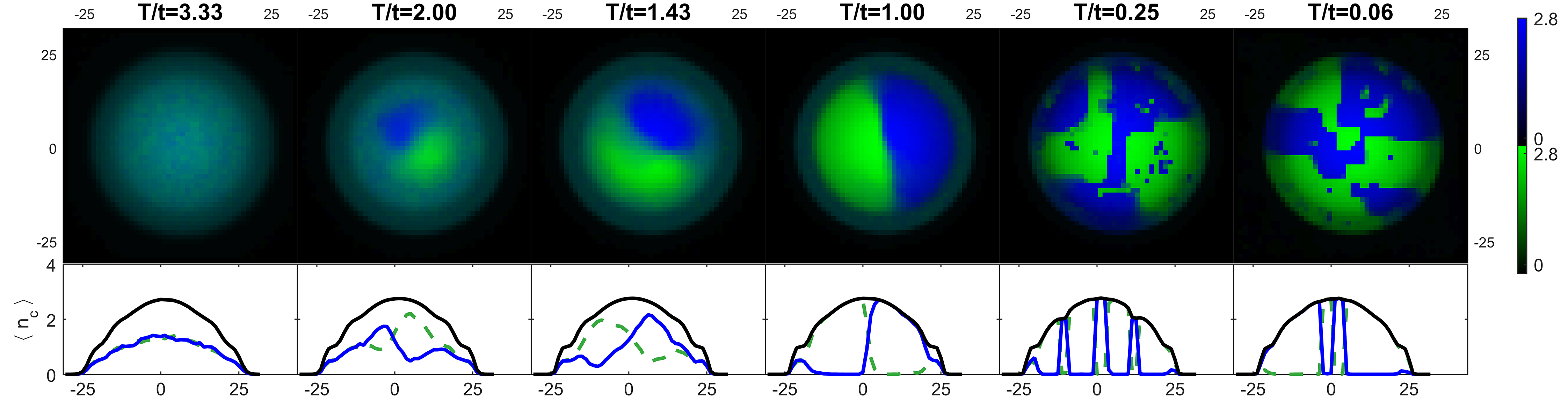}
\caption{Density maps (first row)
    and density profiles (second row) of species A (green) and B (blue), for
    decreasing temperature (left to right) in the SI regime: $U/t=20$,
    $U_{ab}/t=24$ and trapping potential strength $\omega_H/t=0.12$. Density
    profiles $\langle n_c\rangle$ are computed along the
    horizontal axis through the center of the trap. The total density profile is also shown (black-solid line).}
\label{fig:density_prof_hiUW}
\end{figure*}

Spatial shell structures arise also in the SI regime (see
\Fref{fig:density_prof_hiUW}).
For the values $U/t=20$, $U_{ab}/t=24$ considered in \Fref{fig:density_prof_hiUW}, the phase diagram of the homogeneous case predicts
different quantum phases depending on the filling. In particular, at zero temperature and integer
filling, the system is expected to be a dMI. In the trapped case, such a phase
can be observed in the regions of integer filling (density maps at $T/t=0.25$ and $T/t=0.06$ in \Fref{fig:density_prof_hiUW}).
The strong interaction and very low temperatures are responsible for the fragmented structure of these density maps whose metastable character is discussed in the Supplementary information.
Furthermore, at higher temperatures (left columns in
\Fref{fig:density_prof_hiUW}) in the outer regions, we find a mixed phase
forming a thin shell with a $n=1$ plateau, suggesting the presence of a SCF
phase in that region. The SCF at $U_{ab}>U$ is made accessible by temperature
excitations, since the energy separation between the dMI and the SCF is of the
order of $\sim |t^2/U - t^2/U_{ab} + U - U_{ab}|$ \cite{Kuklov2003,Kuklov_errata2015}.

\section*{Experimental realization}
A mixture with $t_a=t_b$ and $U_a=U_b$ can be realized with $^{41}$K atoms in
the two lowest hyperfine states $a=\ket{F=1,m=1}$ and $b=\ket{F=1,m=0}$ (it is
understood that the hyperfine quantum numbers are used only as labels at high
magnetic fields). In presence of a magnetic field $B_0\simeq 675$G the above mixture is predicted to feature
a relatively narrow Feshbach resonance ($\delta B = 0.15$G) between unlike states,
while for like particles the scattering lengths are approximately constant
across the narrow resonance and equal to each other ($a_a\simeq a_b \simeq
60a_0$) \cite{2010-PRA-Lysebo}.  Therefore, with a magnetic field near $B_0$ it
is possible to tune $U_{ab}$, with minimal changes in $U_a$ and $U_b$.
For a heteronuclear mixture of $a=^{41}$K and $b=^{87}$Rb in a square lattice,
tunneling rates can be made nearly equal with an appropriate choice of the lattice step.
For example, for a lattice step $d=380$ nm,
at lattice strengths such as $5\leq U_b/t_b \leq 30$, we have
$0.85\leq t_b/t_a \leq 1.15$ and the ratio $U_b/U_a=0.58$ is constant.

The measurement of parameter $\Delta$  can be obtained directly from high
resolution microscope images \cite{2009-Greiner-microscope,Bloch2010,bakr2015,greiner2015,kuhr2015,cheuk2016}.
However, being a global observable, $\Delta$ does not require knowledge of the local densities
and is also obtained by spectroscopic techniques.
Indeed, the number of sites occupied by both
species can be detected by driving transitions, sufficiently narrow in energy, towards excited states
that can either be internal hyperfine states
\cite{2006-Ketterle-uwaveImaging} or external motional states, such as states of excited lattice bands \cite{2011-Greiner-BandSpectroscopy}.
In practice, the task is eased in proximity of the interspecies Feshbach resonance enhancing $U_{ab}$.
\begin{wrapfigure}{r}{0.4\textwidth}
\begin{center}
  \includegraphics[width=0.4\textwidth]{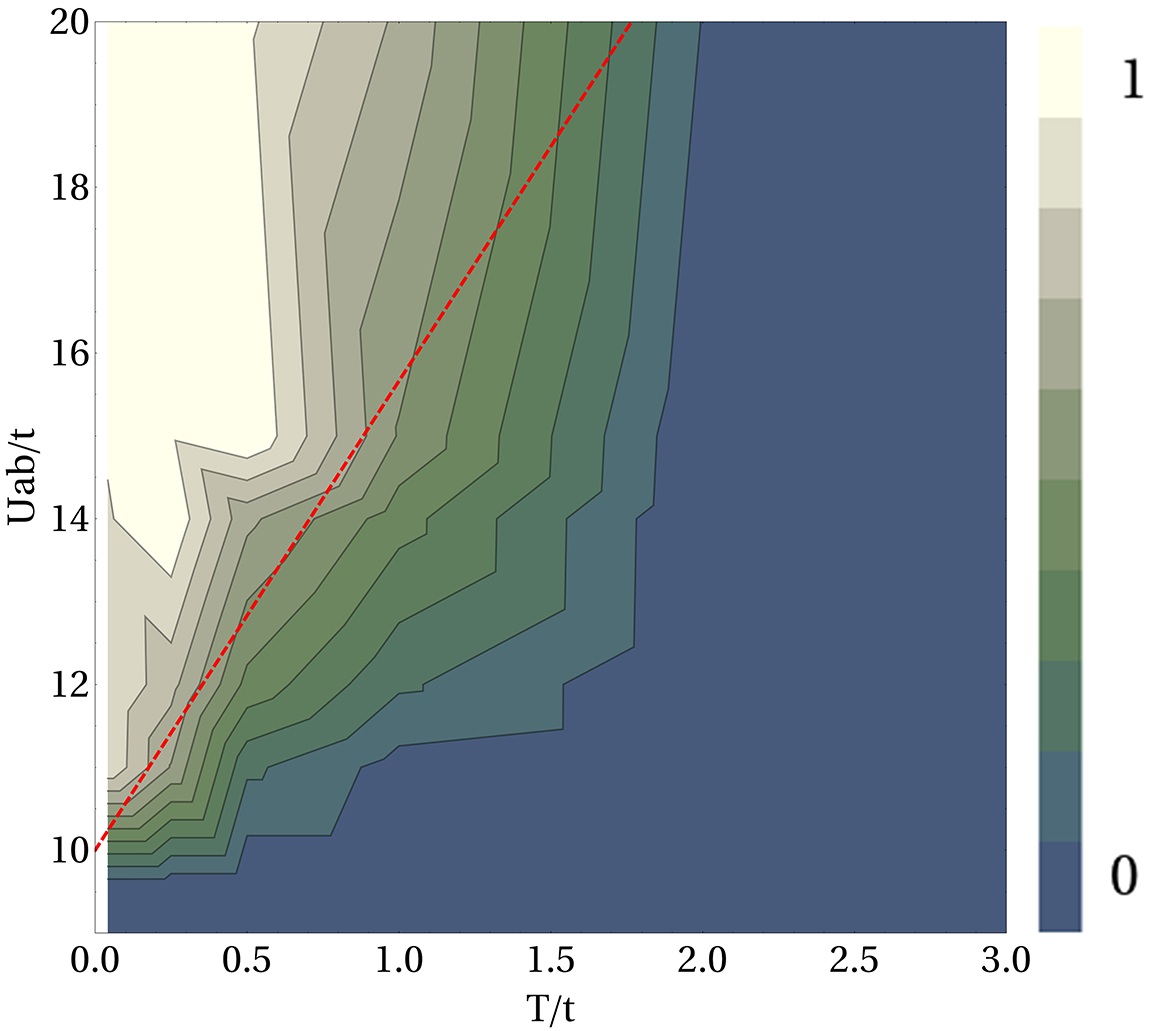}
  \caption{Contour-plot of $\Delta$ as a
    function of temperature $T/t$ and interspecies interactions $U_{ab}/t$, at $U/t=10$. Total filling $n=2$ and linear size of the lattice $L=24$ (same data of Fig. 2, right panel). Red-dashed line shows critical boundary between the demixed and mixed phases computed theoretically through the relation $k_B T= 2(U_{ab} -U)/{\rm ln} (L^4/4)$.}\label{fig:counturPlot}
\end{center}
\end{wrapfigure}
For sake of concreteness, we focus on microwave (or radiofrequency) transitions between internal hyperfine states, and we assume that for excited atoms the interatomic interactions are negligible with respect to $U_{ab}$.
Since $U_{ab}$ is the energy cost required for the formation of a pair $AB$ on the same site, then microwave 
photons will drive transitions in all lattice sites with occupation numbers
$(n_a, n_b)=(\ell, m)$ or $(m, \ell)$ if their frequency is shifted by $\sim - U_{ab} \ell m$ with respect to the bare hyperfine
splitting. Thus, the number of atoms in the excited hyperfine state as a function of the microwave frequency shows several different peaks, each corresponding to a specific pair of occupation numbers, $(\ell, m)$ or $(m,\ell)$. The area of the observed peaks
yields the relative number of the lattice sites, $f(\ell,m)$, with the given fillings.
Once $f(\ell, m)$ values are known for all pairs $(\ell,m)$, the demixing parameter is readily obtained as
$\Delta = \sum_{(\ell, m)} f(\ell,m) (\ell-m)^2 /(\ell+m)^2$.

Finally, we briefly discuss the accuracy of our proposed technique.
For $u\equiv U_{ab}-U \gg t$, i.e. the regime where the above cited spectroscopy is to be performed, $\Delta$
decreases gently with $T/t$, enabling the use of the thermometer in a
continuous fashion. In this range, the value of $\Delta$ is weakly dependent on
$u$, thus the uncertainty on $u$ negligibly affects that on $T/t$.
In addition, since the measured value for $\Delta$ allows to infer $T/t$ values, to convert the temperatures in SI units (i.e. kelvin), we are bound to introduce a relative uncertainty $\delta T/T=\delta t/t$.

The parameters of the Hamiltonian are known with satisfactory precision:
interaction strengths can be determined with relative uncertainty below $10^{-2}$, see e.g.\cite{unc_U}, and similarly the tunneling rates \cite{unc_t}.
Such levels of uncertainty are certainly tolerable for measurements of temperatures in deep optical lattices.

For the sake of clarity, in \Fref{fig:counturPlot}, the right-panel of \Fref{fig:Delta_n1_n2} is represented  in a linear instead of logarithmic scale. This figure shows the contour-plot of $\Delta$ as a function of both $T/t$ and $Uab/t$. One clearly sees how, for sufficiently large values of $U_{ab}/t$, $\Delta$ exhibits a linear behaviour when $T/t$ is varied in the interval $[T_1,T_2]$ with $T_1\approx0.5$ and $T_2\approx2$. In Fig. 6 we also plot the critical boundary (red-dashed line) between the mixed and demixed phases predicted heuristically above ($k_B T= 2(U_{ab} -U)/{\rm ln} (L^4/4)$). We notice that the simulated boundary is very well reproduced by the heuristical formula.

\section*{Conclusions}
We investigated temperature effects on the demixing of a binary
mixture in both the  homogenous and trapped case. Temperature fluctuations progressively destroy spatial separation between the two species with signatures visible also in the momentum distribution. Quite reasonably, higher fillings manifest stronger robustness against temperature-induced mixing. In both the weakly- and the strongly-interacting regime, the demixing parameter $\Delta$ is suppressed in a temperature range of the order of the tunneling energy. We therefore propose to use the experimentally measurable demixing parameter as a thermometer for strongly-correlated binary mixtures in the demixed phase.
With the recent observations of anti-ferromagnetic correlations, reliable thermometry in optical lattices for strongly-interacting regimes is sorely needed to advance the field of `quantum magnetism' with ultracold atoms.

\section*{Methods}

The investigation presented above has been carried out by performing simulations by means of the two-worm algorithm quantum Monte Carlo ~\cite{2worm,prkof_1,prkof_2}, a Path-Integral technique that works within the grand canonical ensemble. It exploits the imaginary-time evolution to evaluate quantum-thermal expectation values of different physical quantities.

\subsection*{The worm algorithm}
According to quantum statistical mechanics the expectation value of a physical observable is given by
\begin{equation}
\langle \hat{O}\rangle = Tr(\hat{\rho} \hat{O})=\sum_\alpha \langle \alpha|\hat{\rho} \hat{O}|\alpha \rangle \label{expO}
\end{equation}
where $\hat{O}$ is the quantum-operator corresponding to the physical observable $O$, $\hat{\rho}=e^{-\beta\hat{H}}/\mathcal{Z}$ is the the density operator and $\mathcal{Z}=Tr(e^{-\beta\hat{H}}) $ the partition function. The parameter $\beta=1/T$ is the inverse temperature and is taken in unit of $k_B=1$. The density operator can be treated as unitary evolution operator in imaginary-time $\tau=i\cdot t$. This allows to estimate the expectation value of a generic observable $\langle \hat{O}\rangle$ as an unitary evolution in imaginary-time between $\tau=0$ and $\tau=\beta$.

Within the interaction picture, the imaginary-time evolution operator takes the form \cite{prkof_2}:
\begin{equation}
e^{-\beta\hat{H}}=e^{-\beta\hat{H_0}}\cdot\mathbf{\hat{T}}e^{-\int_0^\beta\hat{V}(\tau)d\tau}
\end{equation}
where $\mathbf{\hat{T}}$ is the time-ordering operator, and $\hat{V}(\tau)= e^{\hat{H_0}\tau}\hat{V}e^{-\hat{H_0}\tau}$,  while $\hat{H}_0$ and $\hat{V}$ are the diagonal and off-diagonal part of Hamiltonian $\hat{H}$, respectively. The time evolution operator $\hat{\sigma}= \mathbf{\hat{T}}e^{-\int_0^\beta\hat{V}(\tau)d\tau}$ in the interaction picture can be expressed in its iterative expansion form (the Matsubara time-evolution operator)
\begin{equation}
\hat{\sigma}= \mathbf{\hat{T}}e^{-\int_0^\beta\hat{V}d\tau}=
\mathbb{1}  + \hat{\sigma}^{(1)} + \dots + \hat{\sigma}^{(n)}
\end{equation}
where the n-$th$ order term has the form
\begin{equation}
\hat{\sigma}^{(n)}= (-1)^n \int_0^\beta d\tau_n \int_0^{\tau_{n}} d\tau_{n-1} \cdots\int_0^{\tau_{2}} d\tau_1 \hat{V}{(\tau_{n})}\hat{V}{(\tau_{n-1})}\cdots\hat{V}{(\tau)}.
\end{equation}

The chain of operators $\hat{V}{(\tau_{n})}\hat{V}{(\tau_{n-1})}\cdots\hat{V}{(\tau)}$ describes the evolution of the system between $\tau=0$ and $\tau=\beta$. By expanding the off-diagonal operator $\hat{V}$ in an operator basis
\begin{equation}
\hat{V}=\sum_l \hat{K}_l
\end{equation}
such that $\hat{K}_l$ are hermitian and their action on a Fock state of the system results in another state of the same Fock space $\mathcal{H}$
\begin{equation}
\hat{K}_l=\hat{K}_l^\dag,\;\;\;\; \hat{K}_l|\alpha\rangle= k_{l\gamma}|\gamma\rangle,\;\;\;\; with \;\;\;\; |\alpha\rangle,|\gamma\rangle \in \mathcal{H}.
\end{equation}
The n-$th$ order term of the imaginary-time evolution operator can be rewritten in the form
\begin{equation}
\hat{\sigma}^{(n)}= \sum_{l_n, l_{n-1},\dots,l_{1}}(-1)^n \int_0^\beta d\tau_n \int_0^{\tau_{n}} d\tau_{n-1} \cdots\int_0^{\tau_{2}} d\tau_1 \hat{K}_{l_{n}}(\tau_{n})\hat{K}_{l_{n-1}}(\tau_{n-1})\cdots\hat{K}_{l_{1}}(\tau_1).
\end{equation}
It is then possible to rewrite the trace (\ref{expO}) as
\begin{multline}
\langle \hat{O}\rangle = Tr(\hat{\rho} \hat{O})=\frac{1}{\mathcal{Z}}\sum_\alpha \langle \alpha| e^{-\beta\hat{H}_0}\hat{\sigma} \hat{O} |\alpha \rangle =\frac{1}{\mathcal{Z}}\sum_\alpha \langle \alpha| e^{-\beta\hat{H}_0}\hat{\sigma} |\Theta_\alpha \rangle =\\
 =\frac{1}{\mathcal{Z}}\sum_\alpha \sum_n  \sum_{l_n, l_{n-1},\dots,l_{1}}  (-1)^n \int_0^\beta d\tau_n \int_0^{\tau_{n}} d\tau_{n-1} \cdots\int_0^{\tau_{2}} d\tau_1 \langle \alpha| e^{-\beta\hat{H}_0} \hat{K}_{l_{n}}(\tau_{n})\hat{K}_{l_{n-1}}(\tau_{n-1})\cdots\hat{K}_{l_{1}}(\tau_1)|\Theta_\alpha \rangle \label{expO2}
\end{multline}
where $|\Theta_\alpha \rangle=\hat{O} |\alpha \rangle$ is the Fock state resulting from the action of operator $\hat{O}$ on the state $|\alpha\rangle$.
The expectation value of the observable $O$ is then computed as a sum of all the possible evolution in imaginary-time from all the possible initial state $ |\Theta_\alpha \rangle$ at $\tau=0$ to the corresponding definite final state $|\alpha \rangle $ at $\tau=\beta$. The chain of operators $ \hat{K}_{l_{n}}(\tau_{n})\hat{K}_{l_{n-1}}(\tau_{n-1})\cdots\hat{K}_{l_{1}}(\tau_1)$ defines the path in imaginary-time (i.e. worldline) from state $ |\Theta_\alpha \rangle$ to the state $ |\alpha \rangle$. Since the whole computation of $\mathcal{Z}$ and the generation of all the possible paths would be computationally too costly, a Monte Carlo sampling is used to sample only those paths that contribute the most to the expectation value. The Monte Carlo algorithm generates at each Monte Carlo step a different configuration, and,  via a Metropolis Method, accepts or rejects it with a probability that satisfies a proper detailed-balance equation \cite{prkof_1,prkof_2}.

The states of the system described by the two-species Hamiltonian (\ref{H1}) is the tensor product of the Fock states in the spatial mode representation of the two bosonic species
\begin{equation}
|\alpha\rangle=|\alpha_a\rangle\otimes|\alpha_b\rangle=|n_{a1} n_{a2}\dots n_{aM}\rangle\otimes|n_{b1} n_{b2}\dots n_{bM}\rangle
\end{equation}
where $|\alpha_a\rangle$ and $|\alpha_b\rangle$ are the Fock states of species A and B respectively, and $n_{ci}$ the $i$-site occupation number of species $C=A,B$.

The worm algorithm \cite{prkof_1,prkof_2} works in an enlarged configuration space by introducing a disconnected worldline, the worm. This results in working in the grand-canonical ensemble, where particles can be added/removed to/from the system.
``Head'' and ``tail'' of the worm destroy and create a particle in a given site $i$ and imaginary time $\tau$. They correspond to the annihilation operation $c_{i(\tau)}$ and the creation operator $c_{j(\tau)}^\dag$ respectively. Consequently, when the worm is present in the cofiguration
eq. 12 refers to the Green function:
$G(i,j,\tau)=\langle \mathbf{\hat{T}}_\tau c_i(t+\tau)c^\dag_j ( t )\rangle$.

Through operators $\hat{K}_{l}(\tau_m)$, head ($c_{i(\tau)}$), tail ($c_{j(\tau)}^\dag$) the worm moves in space and imaginary time (jump, reconnection, shift in time \dots \cite{prkof_1,prkof_2}) thus generating new configurations.
In our case, in order to explore the configuration space of both the two bosonic species, we use two independent worms that act respectively on the Fock space of the two bosonic species \cite{2worm}.

\subsection*{Estimation of Quantum-Correlators}
To compute the momentum distributions and to check the superfluid/normal-fluid phase transition we estimated quantum-correlators of the form $\langle c_i^\dag c_j\rangle$.
The computation of quantum-correlators $\langle c_i^\dag c_j\rangle$ through the 2-Worm-Algorithm is achieved by collecting statistics of the position of ``head'' and ``tail'' of the worm. Every time the ``tail'' ($c_i^\dag$) and the ``head'' ($c_{j}$) of the worm of species $C=A,B$ are found in position $i$ and $j$ respectively, the correlation-matrix element $CM_{c ij}$ is increased ($c=a,b$).
The quantum-correlators are then estimated as
\begin{equation}
\langle c_i^\dag c_j\rangle \approx \frac{1}{Z_c}CM_{c ij},
\end{equation}
where $Z_c=\sum_{i,j} CM_{c ij}$.

\subsection*{Simulation Setup}
Each simulation is carried out by setting both the temperature and the number of particle.
The worm-algorithm quantum Monte Carlo works within the picture of the grand canonical ensemble at finite temperature. The temperature is controlled through the inverse-temperature parameter $\beta$. The number of particle in the system is controlled via the chemical potentials $\mu_a=\mu_b=\mu$. By carefully tuning the value of $\mu$ it is possible to control the total number of bosons in the lattice.
However, controlling the number of particles of each species turns out to be challenging.
This leads to population densities which varies during the Monte Carlo time.
For example, what typically happens in the regime of the demixing effect is a depletion of one of the bosonic species as the Monte Carlo time goes by \cite{pollet2013}. This situation has to be avoided as in real experimental setup one is capable of controlling independently the number of bosons of each species with a finite precision.
In order to avoid this problem and ensure the conservation of particle of each bosonic species during the simulation we restricted the Hilbert space of the grand canonical ensemble by introducing an upper bound in the quantum fluctuation of the number of particle of each species. This procedure allows us to keep balanced the populations of the two species $N_a\approx N_b\approx N/2$.

\subsection*{Data Availability}
The datasets generated during the current study are available from the corresponding author on reasonable request.

\section*{Acknowledgements}

This work was supported by MIUR (PRIN 2010LLKJBX) and by the NSF (PIF-1552978). The computing for this project was performed at the OU Supercomputing Center for Education and Research (OSCER) at the University of Oklahoma (OU). F.M. acknowledges FP7 Cooperation STREP Project EQuaM (Grant n. 323714).

\section*{Author contributions statement}

F.L. performed the investigation, F.M. conceived the main idea and designed the experimental technique, B.C.S. and V.P. supervised the work. All authors have discussed the results of numerical simulations, and edited and reviewed the manuscript.

\section*{Additional Information}

\textbf{Supplementary information} accompanies this paper at ADRESS.\\
\textbf{Competing financial interests}: The author(s) declare no competing financial interests.

\end{document}


\flushbottom
\maketitle

\subsection*{Spatial-shell structure and momentum distributions}
\begin{figure}[ht!]
  \includegraphics[width=\columnwidth]{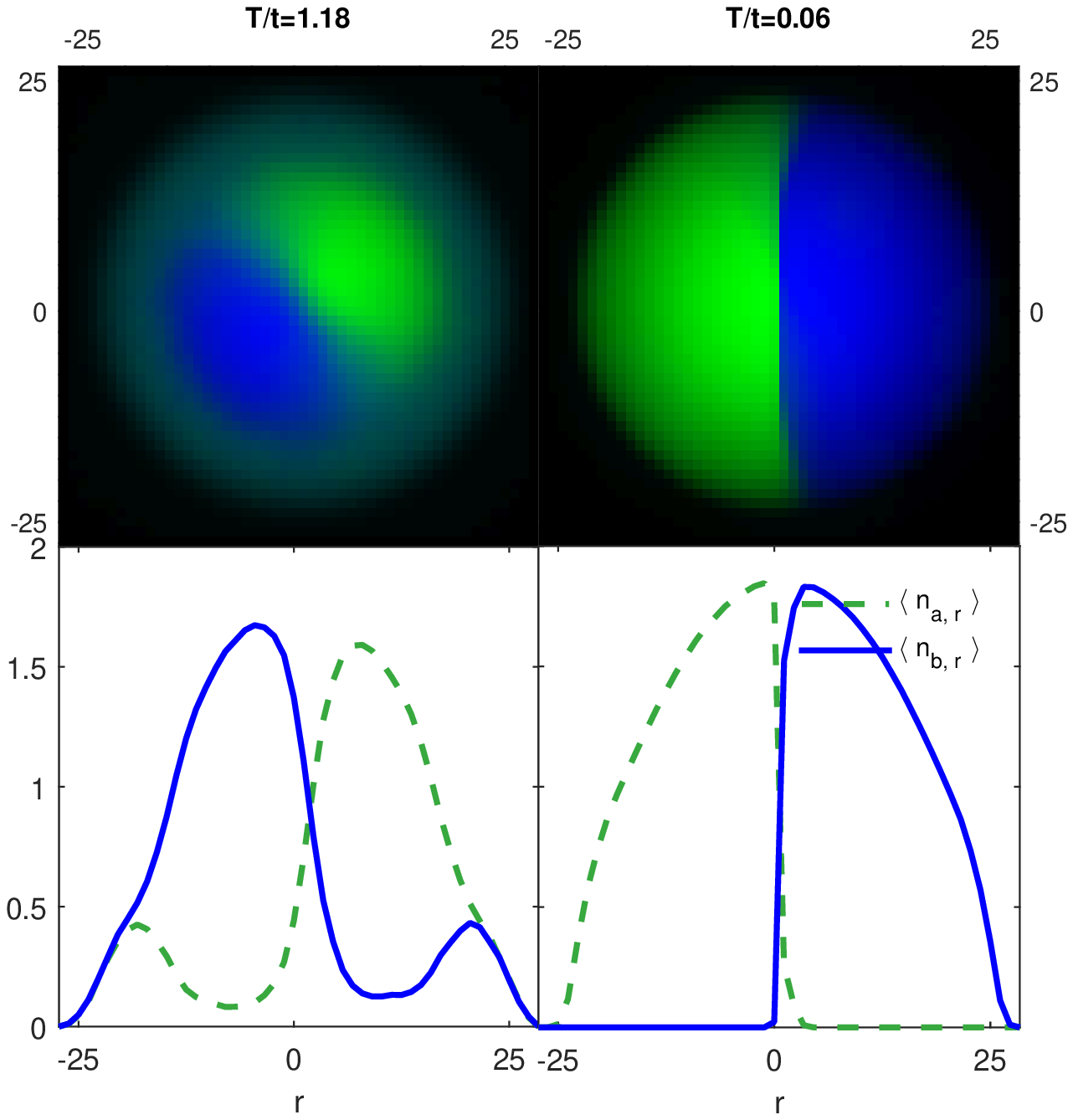}\caption{Density maps (first
    row) and their section at $y=L/2$ along $x$-direction (second row) for
    temperatures $T/t=1.00$ (left column) and $T/t=0.06$ (right column). The
    mixture ($U/t=10$, $U_{ab}/t=15$) is trapped in a harmonic potential of
    strength $\omega_H/t=0.03$. }\label{fig:density_1Dcuts}
\end{figure}

As shown in the main paper, increasing temperatures leads to a progressive
destruction of the spatial demixing between the two species.
This effect first appears in those regions where the entropy associated with demixing
is larger (outer shell where the density is lower and in proximity to the boundary between the two species).
However, it is not restricted only to those region but it extends to the entire lattice. This is shown in the lower row
of Fig. \ref{fig:density_1Dcuts}: in the
case of shell structure (left panel) where the density of one species reaches its maximum, the other does not go identically to zero, as in the
completely demixed case (right panel). A partial but finite mixing of the two
species is present everywhere in the lattice.

\begin{figure}[ht!]
  \includegraphics[width=\columnwidth]{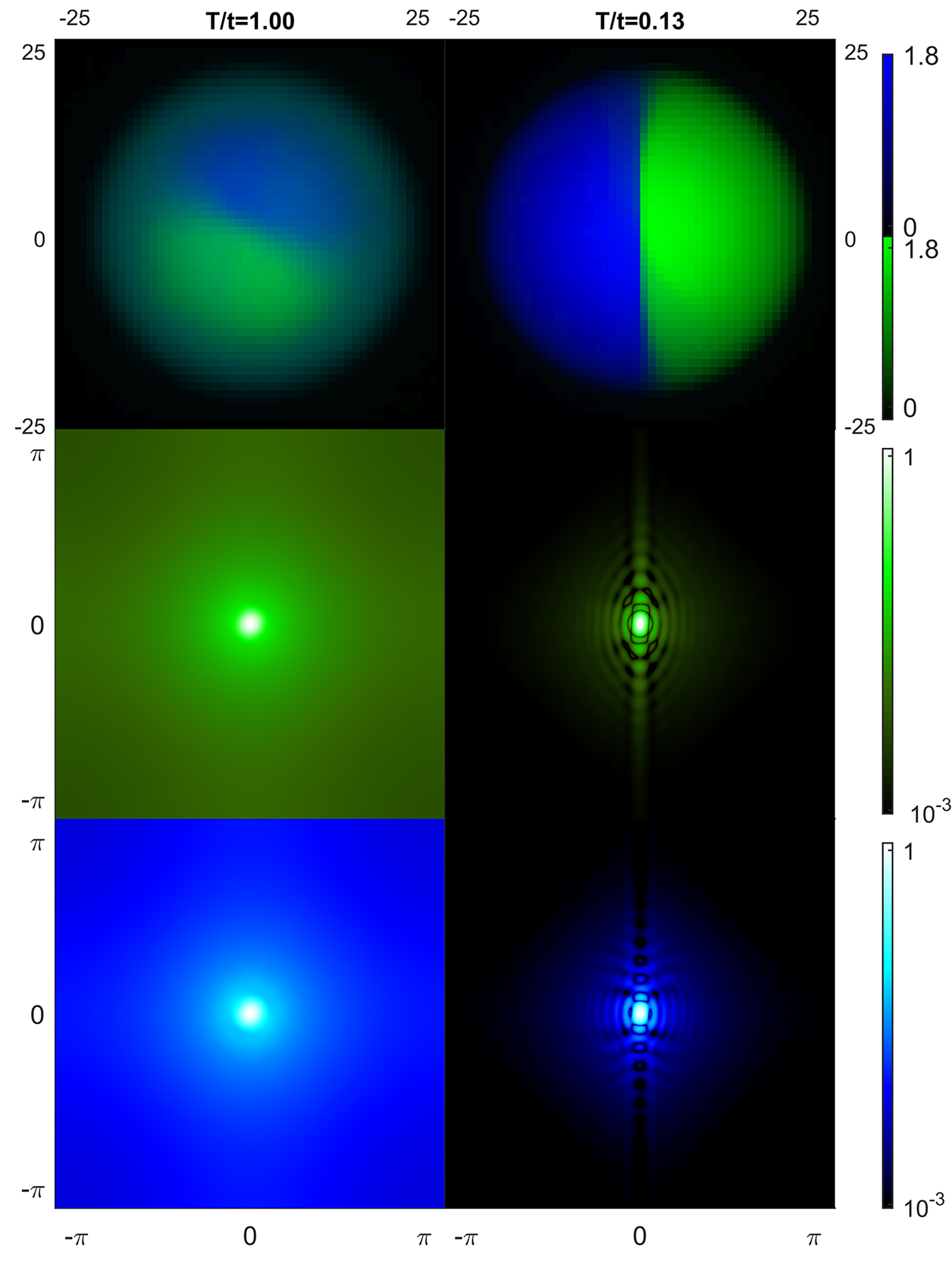}\caption{Density maps (first
    row) and computed momentum distributions for specie A (second-row) and B (third-row) for
    temperatures $T/t=1.00$ (left column) and $T/t=0.06$ (right column). The
    mixture ($U/t=10$, $U_{ab}/t=15$) is trapped in a harmonic potential of
    strength $\omega_H/t=0.03$.}\label{fig:density_TOF_2D_subset}
\end{figure}

As far as the momentum distribution is concerned, by increasing the temperature in the system we observe an abrupt change marking the appearance of the spatial-shell structure and the mixed
phase. This dramatic change is outlined in Fig. \ref{fig:density_TOF_2D_subset},
where we plot together density and momentum distribution of both species for a
spatial-shell structure configuration at $T/t=1.00$ (left column) and a
completely demixed one at $T/t=0.06$ (right column).
At low temperature, we observe oscillations/fringes in the momentum distribution, essentially due to the interference of
waves bouncing from the ``hard-wall'' separating the demixed species.
The oscillations/fringes disappear as soon as the mixing deconfines the two species and phase
coherence is restored among all the lattice sites.

\subsection*{ Momentum distributions of metastable states}
By using the momentum distributions it is possible to
understand if the demixed system is in the ground state configuration or in a
metastable state close to it. In the WI regime, the minimum-energy configuration in a
harmonically-trapped systems features a straight boundary between the
two species \cite{segregPhase1998} (see
Fig. \ref{fig:density_TOF_2D_subset}, right panel). We note, however, that in demixed phases of Bose-Hubbard models
\cite{penna2008,roscilde2007} there exists infinitely-many local minima
featuring metastable states very close in energy to the
ground state. The corresponding configurations of the mean occupation numbers
slightly differ from the ground state and feature, in general, irregular or multiple
boundaries between the two species. Numerical simulations suggest that this kind of configurations
are also observed in the SI regime where, for integer local filling, fragmented dMI regions can be stabilized (see Fig. 5, main text).

\begin{figure}[h!]
\includegraphics[width=\columnwidth]{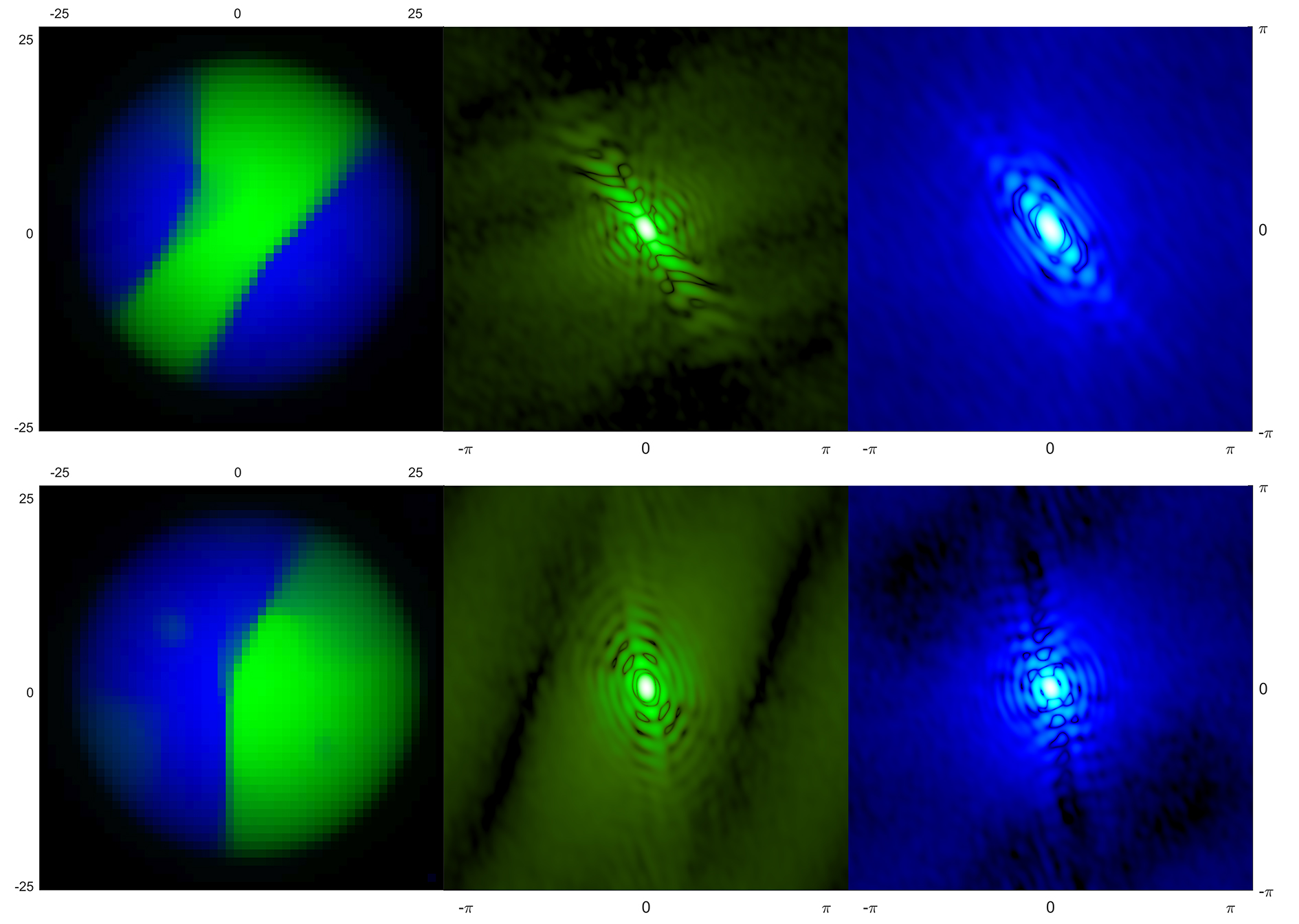}
\caption{Metastable configurations and related momentum distributoins of species A
  (central column) and B (right column). Simulation performed with
  $N\approx1453$ (Upper panel) and $N\approx1644$ (Lower Panel) at $T/t=0.042$,
  $U/t=10$, $U_{ab}/t=15$ and $\omega_H/t=0.03$}\label{fig10}
\end{figure}

By computing the momentum distributions of such
metastable configurations, we find that the irregular/multiple boundaries
exhibited by the demixed states lead to irregular and more complex interference
patterns. Two example of these configurations are shown in
Fig. \ref{fig10}.

\subsection*{Normal fluid Phase Transition}
\begin{figure}[ht!]
  \includegraphics[width=0.6\textwidth]{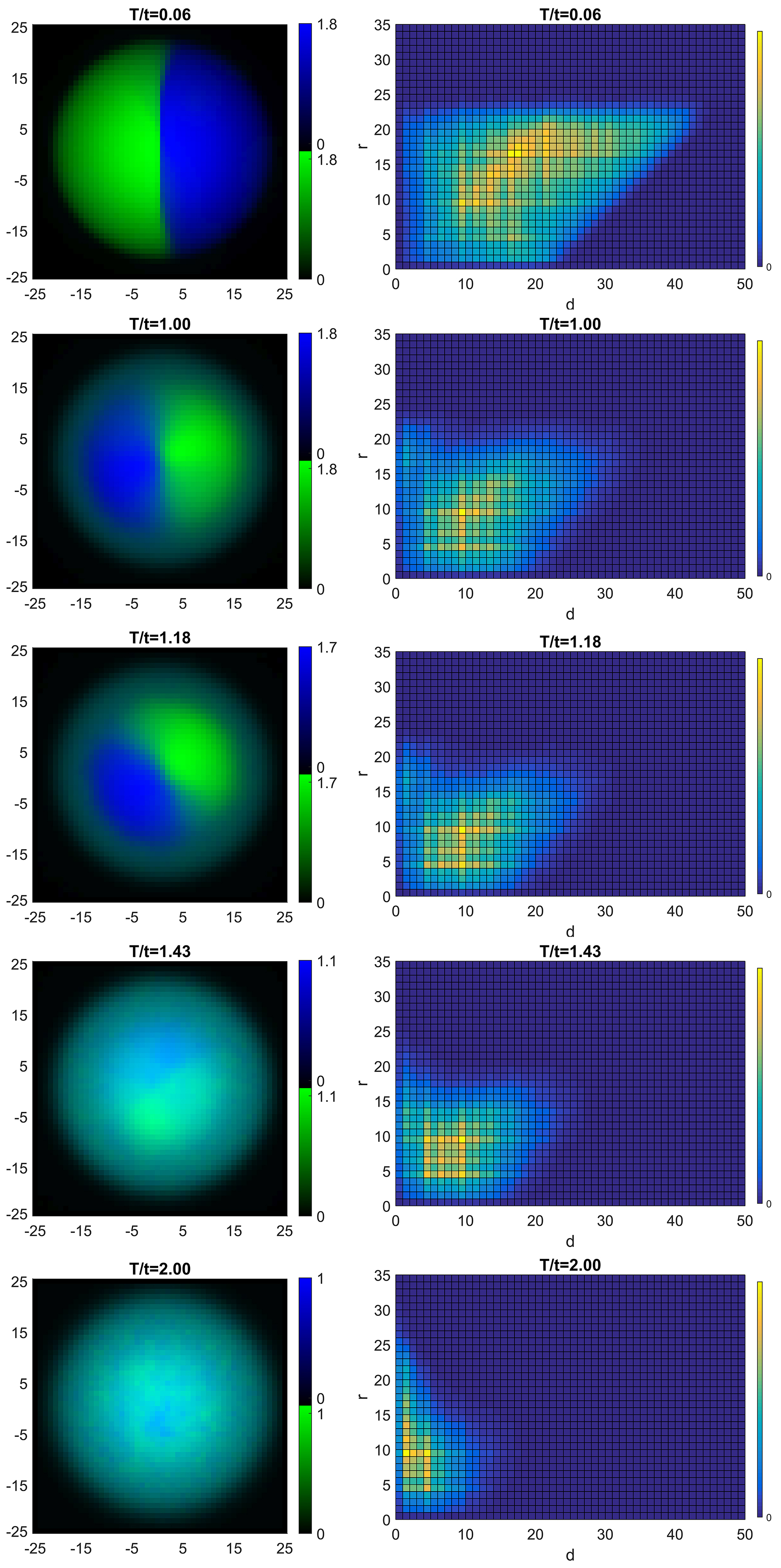}\caption{Field-Field correlators (in arbitrary units) of a single bosonic species for increasing temperatures (right column), and their associated density maps (left column).
  }\label{fig:SFcorr}
\end{figure}

For sufficiently high temperatures, the outer (spatial) shell is filled with a normal fluid (NF).
In order to check the presence of the NF phase we computed the field-field correlators:
\begin{equation}
\langle\Psi_{c}(r)^\dag\Psi_{c}(d)\rangle= \sum_{i\, :\, |\mathbf{r}_i|=r}\,\,\,\,\sum_{j\,:\, |\mathbf{r}_i - \mathbf{r}_j|=d}\langle c^\dag_i c_j \rangle\label{corrSF}
\end{equation}
for species $c=a,b$.
These are obtained from correlators $\langle c^\dag_i c_j \rangle$ by grouping together all the sites $i$ at a given radius $r$ (from the center of the trap) and all the sites $j$ at a given distance $d=|\mathbf{r}_i - \mathbf{r}_j|$ from the sites $i$. In the right column of Fig. \ref{fig:SFcorr} we show the behaviour of correlators (\ref{corrSF}) of species A for increasing temperatures. Due to symmetry, the correlators of species B manifest the same behaviour and therefore are omitted. For the sake of completeness, the left column of Fig. \ref{fig:SFcorr} displays the corresponding density maps.
Fig. \ref{fig:SFcorr} shows that for low temperatures the field-field correlation is spread uniformly all over the occupied lattice sites. On the other hand, we notice that for sufficiently high temperatures the system manifests long-range correlation in a disk-like region around the trap center, and short-range correlation for larger radial distances. In particular, long-range correlations suggest the presence of superfluid phases (dSF and 2SF) in the central disk-like region, while short-range correlations suggest the presence of an external NF shell in the outer portion of the lattice. We notice as well that spatial-shell structures features an increasing thickness of the NF shell for increasing temperatures (last two rows of Fig. \ref{fig:SFcorr}).
